\documentclass[aps,prev,twocolumn,preprintnumbers,floatfix,nofootinbib]{revtex4-1}
\pdfoutput=1

\newcommand{\be}{\begin{equation}}
\newcommand{\ee}{\end{equation}}
\newcommand{\bea}{\begin{eqnarray}}
\newcommand{\eea}{\end{eqnarray}}

\usepackage[utf8]{inputenc}
\usepackage{graphics}
\usepackage{mathrsfs}
\usepackage{amsmath, amsthm, amssymb}
\usepackage{color}
\usepackage{epstopdf}
\usepackage[pdftex]{graphicx}
\usepackage{amssymb}
\usepackage{verbatim}
\usepackage{hyperref}
\usepackage{enumerate}
\usepackage{float}
\usepackage[caption = false]{subfig}
\usepackage[normalem]{ulem}  

\usepackage{amsthm}

\usepackage{ifxetex,ifluatex}
\usepackage{etoolbox}
\usepackage[svgnames]{xcolor}

\usepackage{tikz}

\usepackage{framed}

\newif\ifxetexorluatex
\ifxetex
  \xetexorluatextrue
\else
  \ifluatex
    \xetexorluatextrue
  \else
    \xetexorluatexfalse
  \fi
\fi
\ifxetexorluatex%
  \usepackage{fontspec}
  \usepackage{libertine} 
  \newfontfamily\quotefont[Ligatures=TeX]{Linux Libertine O} 
\else
  \usepackage[utf8]{inputenc}
  \usepackage[T1]{fontenc}
  \usepackage{libertine} 
  \newcommand*\quotefont{\fontfamily{LinuxLibertineT-LF}} 
\fi

\newcommand*\quotesize{60} 
\newcommand*{\openquote}
   {\tikz[remember picture,overlay,xshift=-4ex,yshift=-2.5ex]
   \node (OQ) {\quotefont\fontsize{\quotesize}{\quotesize}\selectfont``};\kern0pt}

\newcommand*{\closequote}[1]
  {\tikz[remember picture,overlay,xshift=4ex,yshift={#1}]
   \node (CQ) {\quotefont\fontsize{\quotesize}{\quotesize}\selectfont''};}

\colorlet{shadecolor}{Azure}

\newcommand*\shadedauthorformat{\emph} 

\newcommand*\authoralign[1]{%
  \if#1l
    \def\authorfill{}\def\quotefill{\hfill}
  \else
    \if#1r
      \def\authorfill{\hfill}\def\quotefill{}
    \else
      \if#1c
        \gdef\authorfill{\hfill}\def\quotefill{\hfill}
      \else\typeout{Invalid option}
      \fi
    \fi
  \fi}
\newenvironment{shadequote}[2][l]%
{\authoralign{#1}
\ifblank{#2}
   {\def\shadequoteauthor{}\def\yshift{-2ex}\def\quotefill{\hfill}}
   {\def\shadequoteauthor{\par\authorfill\shadedauthorformat{#2}}\def\yshift{2ex}}
\begin{snugshade}\begin{quote}\openquote}
{\shadequoteauthor\quotefill\closequote{\yshift}\end{quote}\end{snugshade}}

\begin{document}

\title{The Marshland Conjecture}

\author{David M. C. Marsh}
\email{david.marsh@uni-goettingen.de}
\affiliation{Oskar Klein Centre for Cosmoparticle Physics, Stockholm University, Friedrich-Hund-Platz 1, D-37077 G\"{o}ttingen, Germany} 

\author{J. E. David Marsh}
\email{david.marsh@fysik.su.se}

\vspace{1cm}
\affiliation{Institut f\"{u}r Astrophysik, Georg-August Universit\"{a}t, AlbaNova, Stockholm SE-106 91, Sweden}

\begin{abstract}

We posit the existence of the `Marshland' within string theory. This region is the boundary between the landscape of consistent low-energy limits  of quantum gravity, and the `swampland' of theories that cannot be embedded within string theory because they violate certain trendy and obviously uncontroversial conjectures. 
 The Marshland is probably fractal, and we show some pretty pictures of fractals that will be useful in talks.
We further show that the Marshland contains theories with a large number of light axions, allowing us to cite lots of our own papers. 
We show that the Marshland makes up most of the volume of the landscape, and admits a novel, weakly broken $\mathbb{Z}_2$ `Marshymmetry' that we find strong evidence for by considering a carefully crafted example. 





\end{abstract}

\maketitle

\section{Introduction}

The string theory landscape~\cite{2000JHEP...06..006B,2003dmci.confE..26S} was a really trendy topic about 15 years ago, because it seemed it could have something to say about the cosmological constant problem, and dark energy was also trendy then~\cite{2006IJMPD..15.1753C}. People got pretty bored of it after a while, because it's really hard to do computations in string theory~\cite{1987cup..bookR....G},  and it was generally felt that quantum gravity owes us to be more elegant anyway. 
Recently, however, there has been a surge of renewed interest in landscape-related problems, mostly because  it has been realised that to make citationally rewarding progress, it suffices to boldly \emph{conjecture} that incisive advances have been made. Far-reaching and poorly defined conjectures are particularly useful because they can stimulate vigorous debates about both their assumptions and consequences, and ideally lead to an infinite series of improved conjectures. All the while deferring the doing of serious and consequential work to experimentalists. It is in this spirit that we continue.

%





\section{The Marshland}

The Swampland is defined as the set of apparently consistent low-energy effective field theories that cannot consistently be embedded into quantum gravity\footnote{e.g. string theory. Like most, we couldn't be bothered looking into the others.} \cite{Vafa:2005ui}.
%
%
It has become \emph{really} popular to try to delineate the boundary between the landscape and the swampland, and to speculate about the possible consequences~\cite{1, 2, 3, 4, 5, 6, 7, 8, 9, 10, 11, 12, 13, 14, 15, 16, 17, 18, 19,
20, 21, 22, 23, 24, 25, 26, 27, 28, 29, 30, 31, 32, 33, 34, 35, 36, 
37, 38, 39, 40, 41, 42, 43, 44, 45, 46, 47, 48, 49, 50, 51, 52, 53, 
54, 55, 56, 57, 58, 59, 60, 61, 62, 63, 64, 65, 66, 67, 68, 69, 70, 
71, 72, 73, 74, 75, 76, 77, 78, 79, 80, 81, 82, 83, 84, 85, 86, 87, 
88, 89, 90, 91, 92, 93, 94, 95, 96, 97, 98, 99, 100, 101, 102, 103, 
104, 105, 106, 107, 108, 109, 110, 111, 112, 113, 114, 115, 116, 117, 
118, 119, 120, 121, 122, 123, 124, 125, 126, 127, 128, 129, 130, 131, 
132, 133, 134, 135, 136, 137, 138, 139, 140, 141, 142, 143, 144, 145,  146, 147, 148, 149, 150}. This flurry of activities is giving birth to an emerging field, which can deservedly claim the designation of {\bf Quantum Gravitational Conjecturology} \cite{Buratti:2018onj}.

Strikingly, while the dichotomy between the landscape and the swampland has been widely accepted for more than a decade, few insights from  wetland geomorphology have yet been brought to bear on  quantum gravity. To remedy this issue, and for no other reason,  we Googled the difference between a Marsh and a swamp. 
Swamps typically have trees  and shrubs while Marshes are composed mainly of grasses and reeds.\footnote{Marshes are also known to occur in both Sweden and Great Britain.}  According to a figure we found on Wikipedia \cite{wiki:swamp}, a Marsh can have a swamp to the left and a pretty lake to the right. We then Googled pictures of landscapes and found that lots of them also featured pretty  lakes. It is then natural to assert that: \\

\begin{figure}
\includegraphics[width=0.45\textwidth]{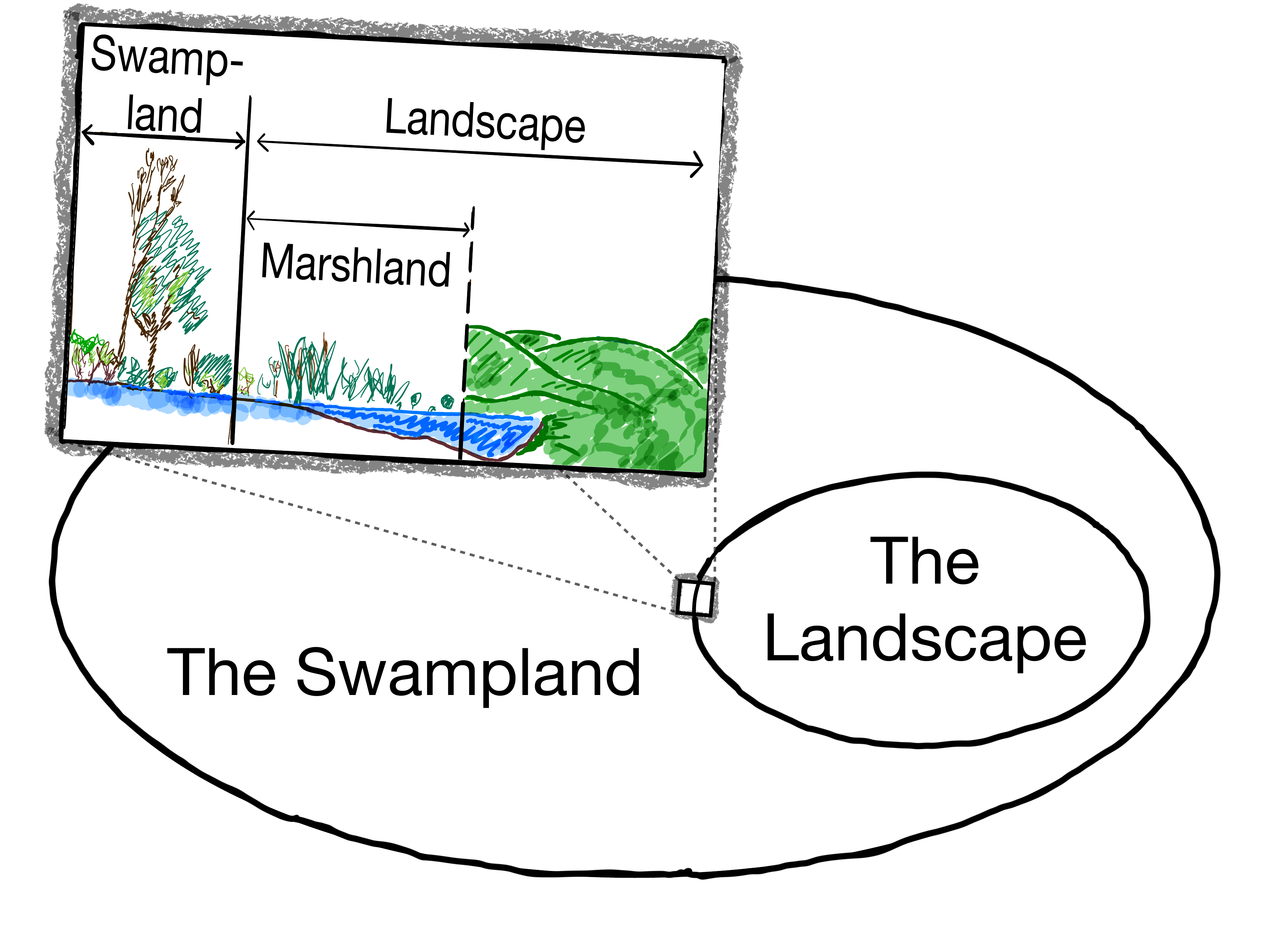}
\caption{The Marshland comprises the  border between the landscape and the swampland.  }
\label{fig:Venn}
\end{figure}

\noindent {\bf The Marshland conjecture:}
The Marshland is the outermost region of the landscape that comprises the
 long-sought 
boundary to the Swampland.  \\ 

Figure \ref{fig:Venn} illustrates our discovery. When we zoom in on the line in the Venn diagram separating the landscape from the Swampland, we can naturally fit in a picture of the Marshland. 

Boundaries are a key part to the AdS/CFT correspondence~\cite{ads_cft}. Our conjecture also contains a boundary, so we think they might be related. 
We're definitely smart enough to write down a metric $g_{ab}$ on the boundary and define conformal transformations, $\Omega$, on it, but we won't. Instead we will just say that it's possible to prove that the Marshland is almost certainly invariant under some group $\mathcal{G}$ with first homotopy group $\pi_1(\mathcal{G})$. We don't use this important result in the rest of the paper, though we will use a corollary without stating it. 
Moreover, 
holographic arguments have demonstrated the existence of a `bootland' --- a map between swampland constraints on the AdS side and bootstrap constraints on the CFT side \cite{Conlon:2018vov}. Eerily, we note that boots are indeed the appropriate footwear in the Marshland.

\begin{figure}
\includegraphics[width=0.4\textwidth]{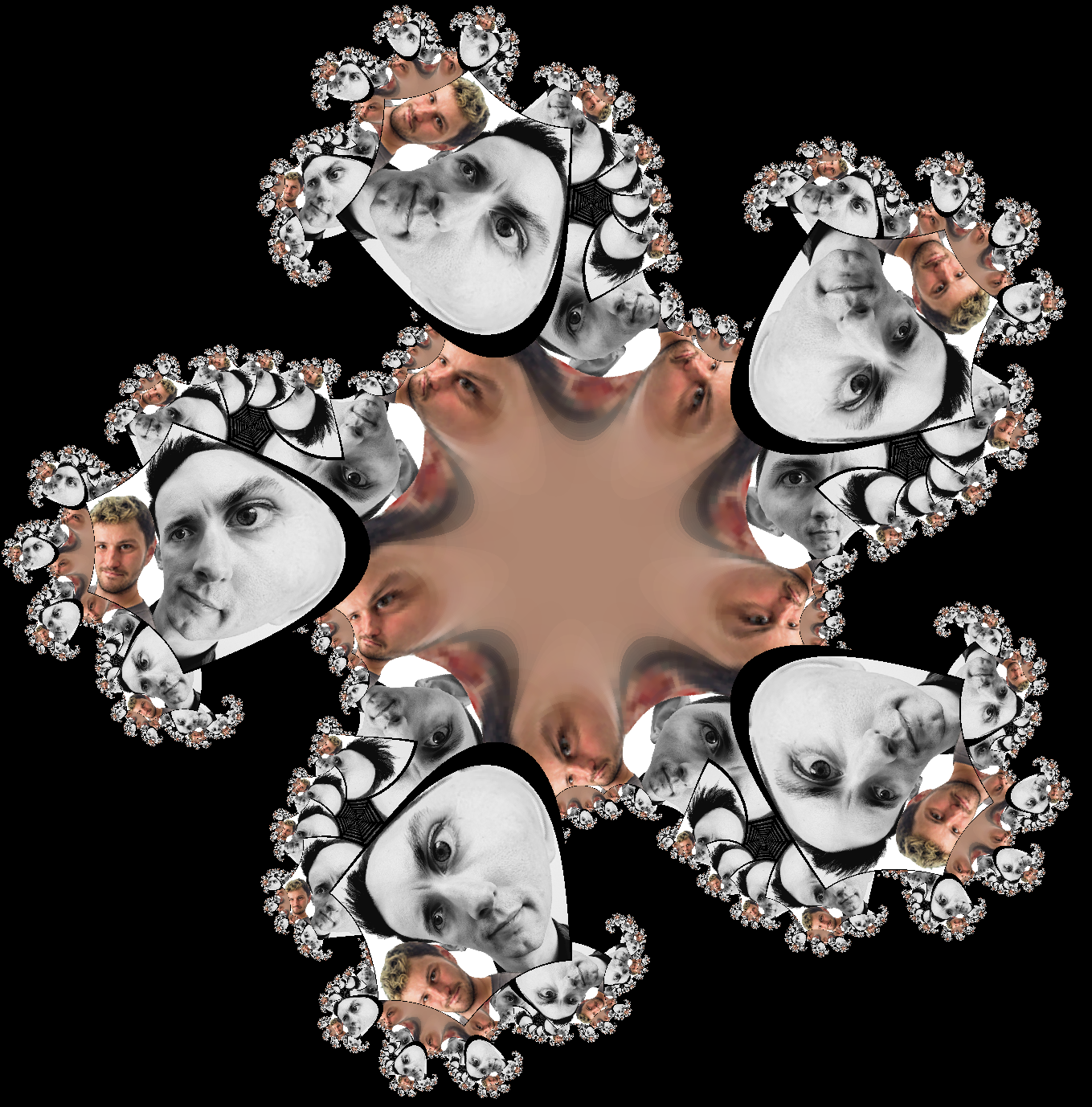}
\caption{Boundaries of complicated things are sometimes fractals, so the Marshland could well be fractal too.}
\label{fig:frac}
\end{figure}

\section{The Marshland is full of axions}

Axions~\cite{pecceiquinn1977,weinberg1978,wilczek1978} are really interesting to everyone~\cite{Marsh:2015xka}, so let's see if we can deduce anything about the relationship of axions to the Marshland.\footnote{There is a vast literature on axions and related topics. For example, see the unbiased selection Refs.~\cite{Marsh:2018zyw,Marsh:2018dlj,Widdicombe:2018oeo,Hlozek:2017zzf,Stott:2017hvl,Fairbairn:2017sil,Abel:2017rtm,Marsh:2017hbv,Stott:2018opm,Armengaud:2017nkf,Diez-Tejedor:2017ivd,Fairbairn:2017dmf,Corasaniti:2016epp,Abazajian:2016yjj,Gonzales-Morales:2016mkl,Helfer:2016ljl,Hlozek:2016lzm,Amendola:2016saw,Marsh:2016vgj,Sarkar:2015dib,Marsh:2015xka,Kim:2015yna,Marsh:2015daa,Marsh:2015wka,Grin:2015bre,Hlozek:2014lca,Fairbairn:2014zta,Bozek:2014uqa,Marsh:2014xoa,Marsh:2014qoa,Iliesiu:2013rqa,Essig:2013lka,Tashiro:2013yea,Marsh:2013ywa,Marsh:2013taa,Amendola:2012ys,Marsh:2012nm,Marsh:2012hjw,Marsh:2011bf,Marsh:2011gr,Marsh:2010wq,Meerburg:2019qqi,Bjorkmo:2019aev,Marsh:2018kub,Bjorkmo:2018txh,Marsh:2018fsu,Bjorkmo:2017nzd,Gallego:2017dvd,Dias:2017gva,Marsh:2017yvc,Cownden:2016hpf,Marsh:2016ynw,Dias:2016slx,Marsh:2015zoa,Brodie:2015kza,Conlon:2015uwa,Marsh:2014nla,Alvarez:2014gua,Marsh:2014gca,Cicoli:2014bfa,Angus:2013sua,Marsh:2013qca,Conlon:2013txa,Conlon:2013isa,Bachlechner:2012at,Berg:2012aq,Marsh:2011aa,Marsh:2011ud,Berg:2010ha,Marsh:2007qp}.} In a generic theory of quantum gravity, the number of axions is constrained during the early Universe due to the gravitational entropy in the vacuum fluctuations of massless fields. Using the entanglement entropy, Conlon computed the bound~\cite{2012JCAP...09..019C}:\footnote{We are a little bit wary of this formula, since it contains the number 13, and when we checked INSPIRE today we found that Edward Witten has exactly 13 ``unknown'' papers~\cite{spires_witten}.}
\be
\sqrt{N}_{\rm ax}\lesssim \frac{\pi\sqrt{13}M_{pl}}{f_a}\, ,
\label{eqn:conlon_eq}
\ee
where $N_{\rm ax}$ is the number of axion fields, $M_{pl}$ is the reduced Planck mass, and $f_a$ is the axion decay constant (here for simplicity assumed to be common to all axions). A theory violating this bound is conjectured to lie in the swampland.  

\begin{figure}
\includegraphics[width=0.5\textwidth]{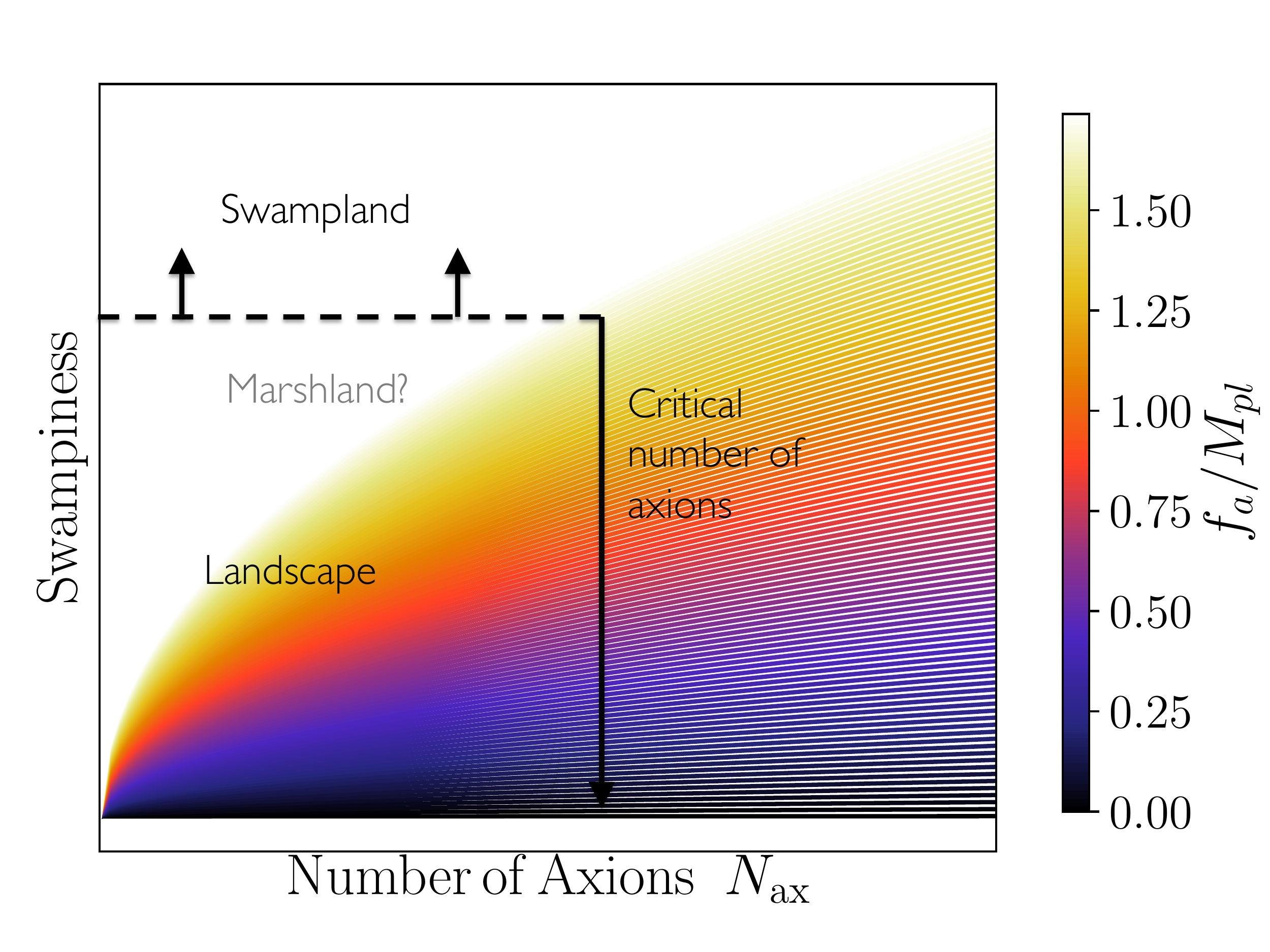}
\caption{The number of light axions, $N_{\rm ax}$, is one parameter that is known to control whether a theory is in the landscape or the swampland. Thus, the theory with the maximum possible number of axions for a given $f_a$ is in the Marshland. This is the critical number of axions, and it is in general a large number.}
\label{fig:n_axions}
\end{figure}

Thus, we conclude that the Marshland should contain theories with the \emph{critical number} of axions, saturating the bound in Eq.~\ref{eqn:conlon_eq}. This idea is illustrated in the schematic Figure \ref{fig:n_axions}. This critical number depends on the value of the decay constant. If $f_a$ also saturates the Weak Gravity Conjecture~\cite{2007JHEP...06..060A}, then $f_a\sim \mathcal{O}(1)M_{pl}$. Taking $f_a = 1.75 M_{pl}$ leads to a prediction for the number of axions in this super-critical part of the Marshland:
\be
N_{\rm ax} = 42\, ,
\ee
which is a pleasing and in no sense contrived result.
\begin{shadequote}[c]{\cite{DouglasAdams}}
``Forty-two'', said Deep Thought, with infinite majesty and calm.
\end{shadequote}

\begin{figure}
\begin{center}
$\begin{array}{@{\hspace{-0.6in}}l@{\hspace{-0.1in}}l}
\includegraphics[width=0.5\textwidth]{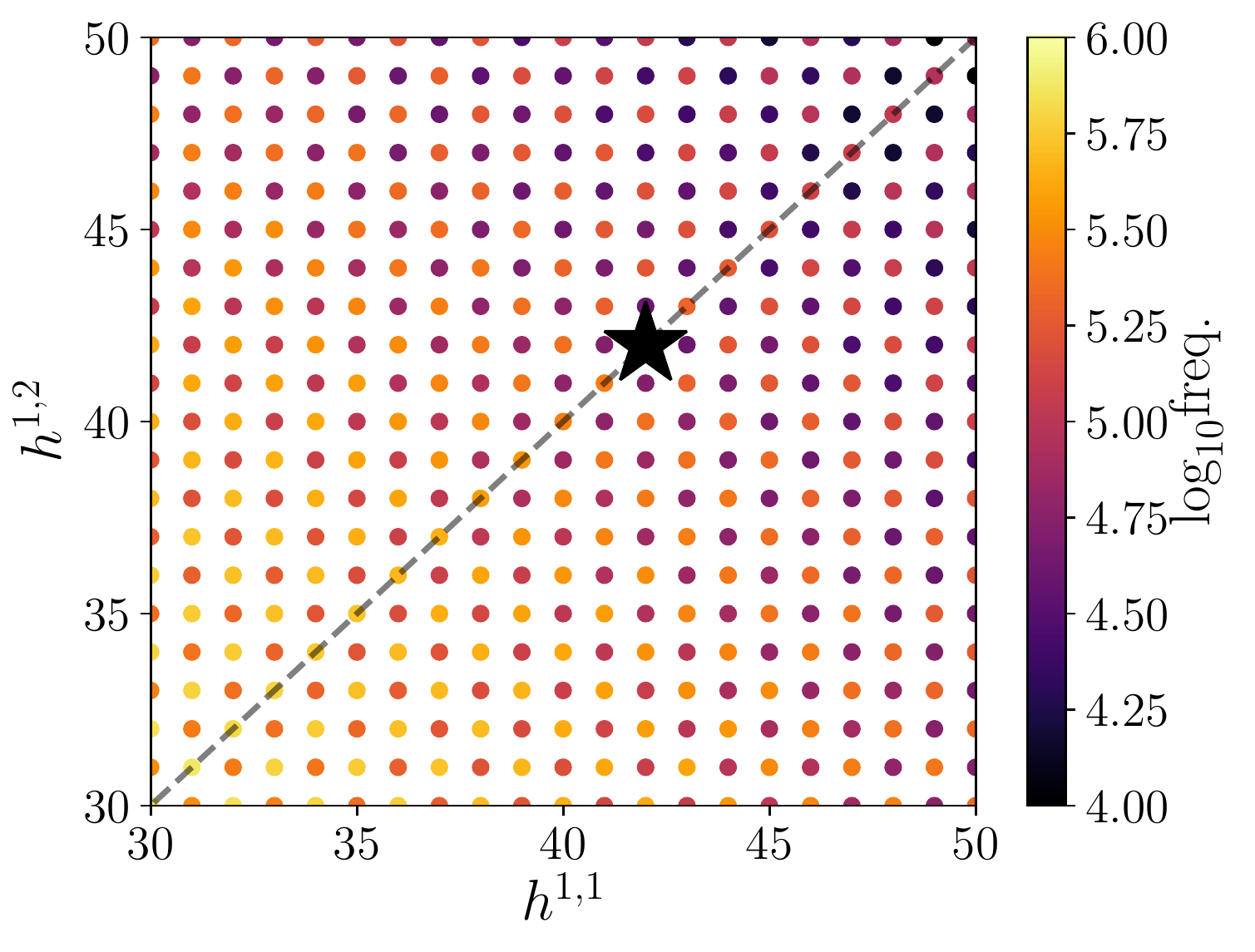}
 \end{array}$
 \caption{Frequency of polytopes in the Kreuzer-Skarke database near $h=42$. The diagonal line marks polytopes which are self-dual under mirror symmetry, and the large star marks the special case $h^{1,1}=h^{1,2}=42$, the self-dual polytopes with the critical number of K\"ahler axions. There are 216523 such polytopes in the database, which is disappointing because 216523 is not a very interesting number.  
 }
\label{fig:KS}
 \end{center}
 
\end{figure}

The Kreuzer-Skarke data base of all reflexive polytopes in four dimensions~\cite{2000math......1106K} can be used to make loads of pretty pictures that may or may not help us visualise aspects of the string landscape. In order to fill some space and show off our skills with \textsc{matplotlib}, we produce one here in Figure~\ref{fig:KS}. The Hodge number $h^{1,1}$ gives the number of (K\"ahler) axions 
in a Type-IIB Calabi-Yau compactification~\cite{2007stmt.book.....B}. Motivated by our previous calculation, we plot the density of polytopes around  $h^{1,2} \approx h^{1,1} \approx42$. It is clear that for any given $h^{1,1}$ there are two populations of polytopes, one with slightly more members than the other. We believe that this anticipates a new approximate symmetry in the landscape, which we will now  discover.  



\section{On the Cosmological Implications of the String Marshland}

We now proceed to make some tenuous predictions that don't quite follow from our conjecture, and that we may be rightly suspected to have wanted to reach all along. 

The Swampland is clearly bigger than the landscape (see Fig.~\ref{fig:Venn}), and is even likely to be of higher dimension than the bulk of the landscape (for example, the parameter direction ``number of Leprechauns'' is not, so far as we know, a direction in the landscape, but could well be one in the swampland). The Marshland, being the boundary region, will then have a dimension in-between that of the bulk of the landscape and the Swampland --- the Marshland is fractal (as already demonstrated in Figure~\ref{fig:frac}. Take another look: isn't it great?). We expect  that most of the volume of the landscape is in the boundary Marshland, but precision studies of the geometry of the landscape Venn diagram will be required to fully settle this question.  




%


Continuous global symmetries are believed to be forbidden in the whole landscape, including the Marshland. But this simply means that the Marshland should contain an approximate $\mathbb{Z}_2$ symmetry that acts non-linearly on world-lines of certain particularly marshy objects. 
This predicted `Marshymmetry' is probably emergent from deep string theory consistency relations, because that sounds pretty cool. 

\begin{figure}
\includegraphics[width=0.45\textwidth]{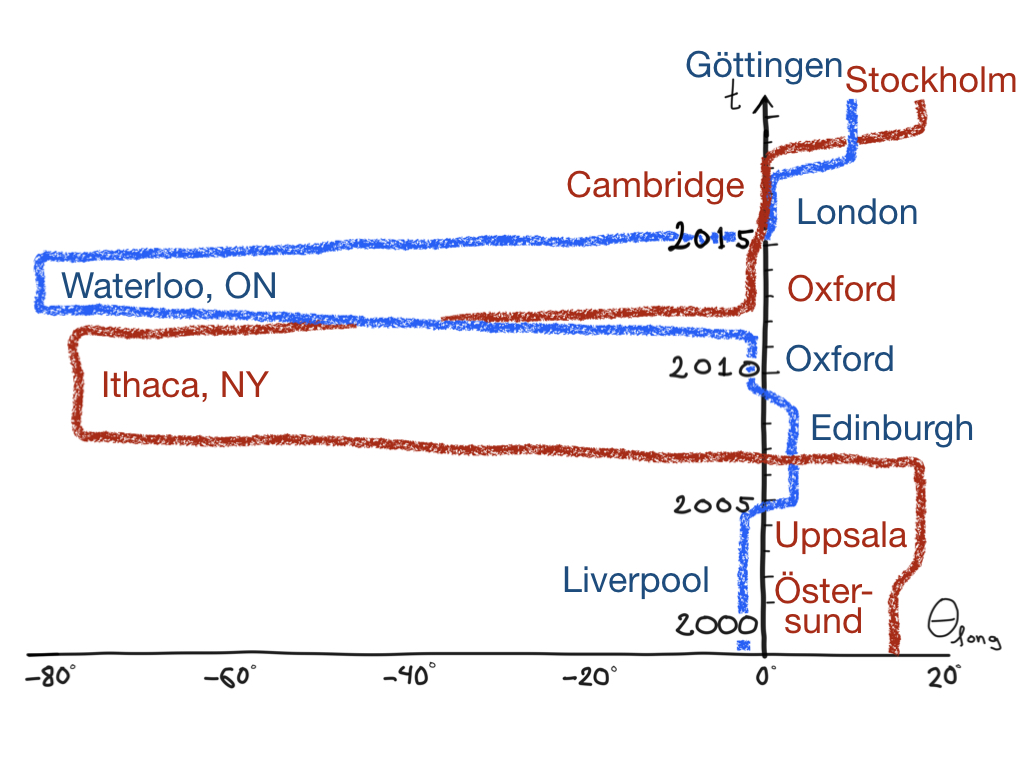}
\caption{The approximate $\mathbb{Z}_2$ Marshymmetry may well act non-linearly to interchange the authors' world-lines, here plotted against longitudinal angle.}
\label{fig:worldline}
\end{figure}

Remarkably, hints of the Marshymmetry  may be evident in the braided world-lines of the present authors, cf.~Figure \ref{fig:worldline}. 
Without the Marshymmetry, an unsettlingly large amount of fine-tuning or a staggering random fluke would be required to explain how two physicists sharing both given and last names --- the latter itself foretelling the Marshland discovery --- would graduate the same year, work on overlapping topics (including axions, cosmology, and the landscape, see again Refs.~\cite{Marsh:2018zyw,Marsh:2018dlj,Widdicombe:2018oeo,Hlozek:2017zzf,Stott:2017hvl,Fairbairn:2017sil,Abel:2017rtm,Marsh:2017hbv,Stott:2018opm,Armengaud:2017nkf,Diez-Tejedor:2017ivd,Fairbairn:2017dmf,Corasaniti:2016epp,Abazajian:2016yjj,Gonzales-Morales:2016mkl,Helfer:2016ljl,Hlozek:2016lzm,Amendola:2016saw,Marsh:2016vgj,Sarkar:2015dib,Marsh:2015xka,Kim:2015yna,Marsh:2015daa,Marsh:2015wka,Grin:2015bre,Hlozek:2014lca,Fairbairn:2014zta,Bozek:2014uqa,Marsh:2014xoa,Marsh:2014qoa,Iliesiu:2013rqa,Essig:2013lka,Tashiro:2013yea,Marsh:2013ywa,Marsh:2013taa,Amendola:2012ys,Marsh:2012nm,Marsh:2012hjw,Marsh:2011bf,Marsh:2011gr,Marsh:2010wq,Meerburg:2019qqi,Bjorkmo:2019aev,Marsh:2018kub,Bjorkmo:2018txh,Marsh:2018fsu,Bjorkmo:2017nzd,Gallego:2017dvd,Dias:2017gva,Marsh:2017yvc,Cownden:2016hpf,Marsh:2016ynw,Dias:2016slx,Marsh:2015zoa,Brodie:2015kza,Conlon:2015uwa,Marsh:2014nla,Alvarez:2014gua,Marsh:2014gca,Cicoli:2014bfa,Angus:2013sua,Marsh:2013qca,Conlon:2013txa,Conlon:2013isa,Bachlechner:2012at,Berg:2012aq,Marsh:2011aa,Marsh:2011ud,Berg:2010ha,Marsh:2007qp}) at sometimes the same institution (at different times), and even co-author a paper heralding a new era in quantum gravitational conjecturology.  We interpret this carefully selected example as strong evidence for the Marshland conjecture.

\section{Conclusions}
\begin{shadequote}[c]{Slartibartfast, \cite{DouglasAdams}}
[T]he chances of finding out what really is going on are so absurdly remote that the only thing to do is to say ``Hang the sense of it'' and just keep yourself occupied. Science has achieved some wonderful things, I know, but I'd far rather be happy than right any day.
\end{shadequote}

Despite the heroic efforts in the attempts to conjecture the deep secrets of quantum gravity, the origin of the neutrino masses, the solution to the cosmological constant problem, and the meaning of life, it has generally been felt that something is still lacking. We have shown that the answer is in the Marshland. In an infinite Universe everything that is possible, no matter how rare, happens an infinite number of times\footnote{This is a bit of a counter example to the need for the $\mathbf{Z}_2$ symmetry postulated earlier, but we're working to a deadline for this joke to work and are past caring.}. In an infinite theory space, similar things happen. For example, there are infinite variations of this paper posted to the arXiv on days other than April 1st.


\acknowledgements{M. David J. Marsh and E. David C. Marsh acknowledge the support given to us by many accidental citations.
No blame for this paper should fall on Francesca Day, Blake Sherwin or Prateek Agrawal. They meant no harm. }
%
\bibliographystyle{h-physrev3}

\bibliography{axion,marsh,swamp}

\end{document}